\documentstyle[psfig]{espart}
\begin{document}
\begin{frontmatter}
  \title{{\small J. Stat. Phys., Vol.80, 1443-1452 (1995)}\\[1.5cm]
Guessing probability distributions from small samples} 
  \author[HUB,CHIC]{Thorsten P\"oschel\thanksref{bylinetp}} 
  \author[HUB]{Werner Ebeling\thanksref{bylinewe}} 
  \author[HUB]{Helge Ros\'e\thanksref{bylinehr}} 
  \address[HUB]{Humboldt--Universit\"at zu
    Berlin, Institut f\"ur Physik, Unter den Linden 6, D--10099 
    Berlin, Germany, 
    http://summa.physik.hu-berlin.de:80/$\sim$thorsten/}
  \address[CHIC]{The James Franck Institute, The University of Chicago, 
    5640 South Ellis Av., Chicago, Illinois 60637}
  \thanks[bylinetp]{E--mail: thorsten@hlrsun.hlrz.kfa--juelich.de}
  \thanks[bylinewe]{E--mail: werner@itp02.physik.hu--berlin.de}
  \thanks[bylinehr]{E--mail: rose@summa.physik.hu--berlin.de}
\begin{abstract}
We propose a new method for the calculation of the statistical
properties, as e.g. the entropy, of unknown generators of symbolic
sequences. The probability distribution $p(k)$ of the elements $k$ of
a population can be approximated by the frequencies $f(k)$ of a sample
provided the sample is long enough so that each element $k$ occurs
many times. Our method yields an approximation if this precondition
does not hold. For a given $f(k)$ we recalculate the Zipf--ordered
probability distribution by optimization of the parameters of a
guessed distribution. We demonstrate that our method yields reliable
results.
\end{abstract}
\end{frontmatter}
\section{Introduction}
Given a statistical population of discrete events $k$ generated by a
stationary dynamic process, one of the most interesting statistical
properties of the population and hence of the process is its entropy. If
the sample space, i.e. the number of different elements which are
allowed to occur in the population, is small compared with the size of
a drawn sample one can approximate the probabilities $p(k)$ of the
elements $k$ by their relative frequencies $f(k)$ and one finds for
the observed entropy $H_{obs}$
\begin{equation}
H=-\sum_k p(k) \log p(k) \approx -\sum_k f(k) \log f(k)=H_{obs}~.
\label{approximation}
\end{equation}
If the number of the allowed different events is not small compared
with the size of the sample the approximation $p(k)\approx f(k)$
yields dramatically wrong results. In this case the knowledge of the
frequencies is not sufficient to determine the entropy. The aim of
this paper is to provide a method to calculate the entropy and other
statistical characteristics for the case that the approximation
(\ref{approximation}) does not hold.

\begin{figure}[ht]
\centerline{\psfig{figure=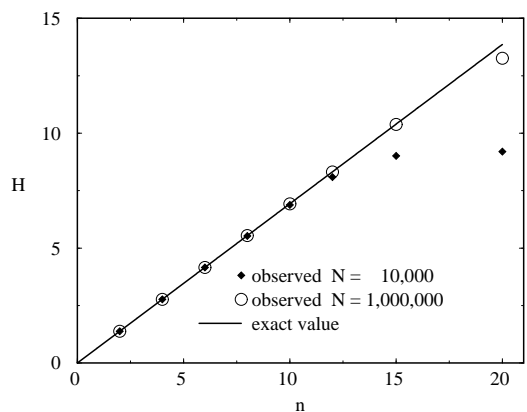,width=7cm,angle=270}}
\caption{The observed entropy for $N$ words of length $n$ from a
Bernoulli sequence.}
\label{hbern}
\end{figure}

An interesting example of such systems are subsequences (words) of
length $n$ of symbolic sequences of length $L$ written using an
alphabet of $\lambda$ letters. Examples are biosequences like DNA
($\lambda=4$, $L {<\atop\approx} 10^9$), literary texts ($\lambda
\approx 80$ letters and punctuation marks, $L {<\atop\approx} 10^7$)
and computer files ($\lambda=2$, $L$ arbitrary). For the case of
biosequences there is a variety of $\lambda^n=1,048,576$ different
words of length $n=10$. To measure the probability distribution of the
words directly by counting their frequencies we need at least a
sequence of length $10^8$ to have reliable statistics. Therefore the
ensemble of subsequences of length $n$ is a typical example where the
precondition does not hold. To illustrate the problem we calculate the
observed entropy $H_{obs}^{(n)}$ for $N$ words of length $n$ in a
Bernoulli sequence with $\lambda=2$ where both symbols occur with the
same probability. The exact result is $H^{(n)}=n\log\lambda$. In
figure~\ref{hbern} we have drawn the values of $H^{(n)}$ and
$H_{obs}^{(n)}$ over $n$. The observed entropy values are correct for
small word length $n$ when we can approximate the probabilities by the
relative frequencies. For larger word length, however, the observed
entropies are significantly below the exact values, even for very
large samples (circles: $N=10^6$, diamonds: $N=10^4$).

Under several strong preconditions the probabilities of words in
sequences can be estimated from the frequencies using various
correction methods~\cite{justice,herzel,schmitt}. The advanced
algorithm proposed in~\cite{schmitt} is based on a theorem by McMillan
and Khinchin \cite{mcmillan} saying that for word length $n
\rightarrow \infty$ the frequencies of the admitted substrings of a
sequence are equally distributed. If one is interested in the
entropies for {\it finite} words, however, the theoretical basis to
apply this theorem is weak and there is no evidence about the
reliability of the results. Moreover this theorem is proven for Markov
sequences only. In sequences gathered from natural languages,
biosequences and other natural or artificial sources it is very
unlikely that the probabilities of the words of interesting length,
e.g.~words or sentences for languages, amino acids or elements of the
hidden ``DNA language'' for biosequences, are equally
distributed. Otherwise we had to assume that all English
five--letter--words are equally frequent. Certainly this is not the
case.

\section{Description of the method}
To calculate the entropy of a distribution it is not necessary to
determine for each event $k$ the probability $p(k)$. It is sufficient
to determine the values of the probabilities without knowing which
probability belongs to which event. Generally spoken if we assume to
have $K$ events there are $K!$ different relations $k\leftrightarrow
p$. We need not to determine one particular (the correct relation) but
only one arbitrary of them. Hence the calculation of the entropy is
$K!$ times easier than to determine the probability $p(k)$ for each
event $k$. We assume a special order where the first element has the
largest probability, the second one the second largest etc.  We call
this distribution Zipf--ordered. Zipf ordering means that the
probabilities of the elements are ordered according to their rank and
therefore the distribution $p(k)$ is a monotonically decaying
function. The following procedure describes a method how to
reconstruct the Zipf--ordered probability distribution $p(k)$ from a
finite sample.

Provided we have some reason to expect (to guess) the parametric form
of the probability distribution. As an example we use a simple
distribution $p(k,\alpha,\beta,\gamma)$ with $k=1,2,\dots$ consisting
of a linearly decreasing and a constant part
\begin{equation}
p(k)=\left\{ 
  \begin{array}{r r l}
  \frac{2-\alpha\,\beta}{2\,\gamma}+\alpha\,(1-\frac{k}{\beta}) & 
   : &  1 \le k < \beta\\
  (2-\alpha\,\beta)/(2\,\gamma) &:& \beta \le k \le \gamma \\ 
   0                            &:& k > \gamma {\hspace{0.3cm} .}
  \end{array} \right.
\label{probability}
\end{equation}
Then the algorithm runs as follows:
\renewcommand{\labelenumi}{\roman{enumi}.}
\begin{enumerate}
\item Find the frequencies $F(k)$ for the $N$ events $k$ and order
  them according to their value (Zipf--order). The index $k$ runs over
  all {\it different} events occurring in the sample ($k \in \{1 \dots
  K^{MAX}$\}). Note: there are $N$ events but only $K^{MAX}$ different
  ones. Normalize this distribution
    $F_{1}(k)=F(k)/N$.
  There are various sophisticated algorithms to find the frequencies
  of large samples and to order them (e.g.~\cite{knuth}). As in
  earlier papers \cite{epa} we applied for finding the elements a
  ``hashing''--method and for sorting a mixed algorithm consisting of
  ``Quicksort'' for the frequent elements and ``Distribution
  Counting'' for the long tail of elements with low frequencies.
\item Guess initial conditions for the parameters (in our case $\alpha$,
  $\beta$ and $\gamma$).
\item Generate $M$ samples of $N$ random integers ($RI_k^m$,
$k=1\dots N$, $m=1\dots M$) according
  to the parametric probability distribution
  $p(k,\alpha,\beta,\gamma)$. In the following examples we used
  $M=20$. Order each of the samples according to the ranks
  $f_i(k,\alpha,\beta,\gamma)$ ($i=1\dots M$).
  Average over the $M$ ordered samples
  \begin{equation}
  \overline{f(k,\alpha,\beta,\gamma)} = \frac{1}{M}\sum_{i=1}^M
    f_i(k,\alpha,\beta,\gamma)
  \end{equation}
  with $k \in \left\{1, k^{max} \right\}$ and $k^{max} = \max\,\left(
  k_i^{max}, \,\,\, (i=1\dots M)\right)$.
  Since we want to determine the averaged or typical Zipf--ordered
distribution, it is important to order the elements first and then to average.
  Normalize the averaged distribution of the frequencies
  \begin{equation}
  \overline{f_{1}(k,\alpha,\beta,\gamma)}=
  \left(\sum_{k=0}^{k^{max}}\overline{f(k,\alpha,\beta,\gamma)}
  \right)^{-1}~\overline{f(k,\alpha,\beta,\gamma)}~.
  \end{equation}
\item Measure the deviation $D$ between the normalized averaged
  simulated frequency distribution
  $\overline{f_{1}(k,\alpha,\beta,\gamma)}$ and the frequency
  distribution $F_{1}(k)$ of the given sample according to a certain
rule, e.g.
  \begin{equation}
    D = \sum_{k=1}^{K}
    \left(\frac{\overline{f_{1}(k,\alpha,\beta,\gamma)}}{F_{1}(k)}-1 \right)^2 
    ~\mbox{,}~~
    K = \max \left \{k^{max}, K^{MAX} \right\}~. 
  \end{equation}
\item Change the parameters of the guessed probability distribution
  $p(k)$ (in our case the parameters $\alpha$, $\beta$ and $\gamma$)
  due to an optimization rule (e.g.~\cite{recipes}) which minimizes
  $D$ and proceed with the third step until the deviation $D$ is
  sufficiently small.
\item Extract the interesting statistical properties out of the
  probability distribution $p(k)$ using the parameters $\alpha^*$,
  $\beta^*$ and $\gamma^*$ which have been gathered during the
  optimization process.
\end{enumerate}

\section{Examples}
\subsection{Entropy of artificial sequences}
\begin{figure}[ht]
\centerline{\psfig{figure=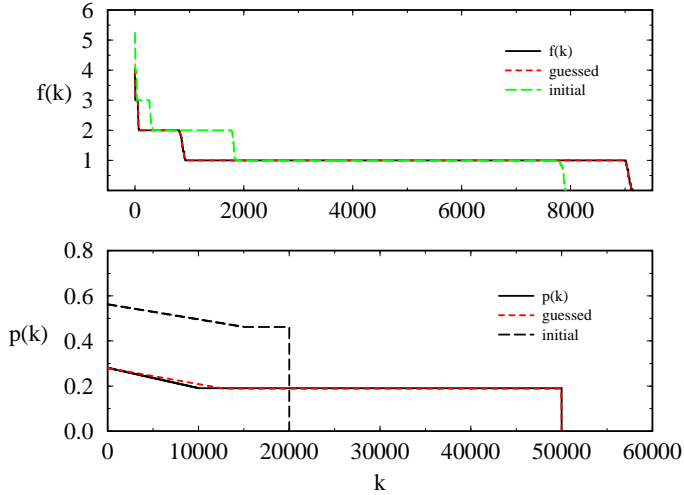,width=9cm,angle=270}}
\caption{The probability distribution $p(k)$ (eq.~\ref{probability})
  and the Zipf--ordered frequencies $f(k)$ corresponding to this
  distribution. The dashed lines which can almost not be distinguished
  from the solid lines display the guessed distributions. The initial
  distributions before optimization is drawn with wide dashes.}
\label{test}
\end{figure}

We generated a statistical ensemble $N=10^4$ according to the
probability distribution eq.~(\ref{probability}) with
$\alpha=9.0\cdot10^{-6}$, $\beta=10,000$ and
$\gamma=50,000$. Fig.~\ref{test} (solid lines) shows the probability
distribution $p(k)$ and the Zipf--ordered frequencies $f(k)$.

Optimizing the parametric guessed probability distribution
using the proposed method we find for the optimized
parameters $\alpha^*=9.22\cdot10^{-6}$, $\beta^*=12,900$ and
$\gamma^*=50,000$, i.e. the guessed and the actual distributions fall
almost together. 
Since we know the original probability distribution
(eq.~\ref{probability}) we can compare its exact entropy with the
entropy of the guessed probability distribution $H_{guess}$ and with
the observed entropy $H_{obs}$ due to
eqs.~(\ref{hdefinition1},\ref{hdefinition}).
\begin{eqnarray}
  &&H_{guess}(k,\alpha^*,\beta^*,\gamma^*) =
     -\sum p(k,\alpha^*,\beta^*,\gamma^*)\cdot 
           \log p(k,\alpha^*,\beta^*,\gamma^*)
  \label{hdefinition1}\\
  &&H_{obs}(k)=-\sum F_{1}(k)\cdot
  \log F_{1}(k)
  \label{hdefinition} 
\end{eqnarray}
We found $H_{obs}=9.0811$ and $H_{guess}=10.8147$, the exact value
according to $p(k,\alpha,\beta,\gamma)$ (eq.~\ref{probability})
is $H=10.8188$.

\begin{figure}[ht]
\centerline{\psfig{figure=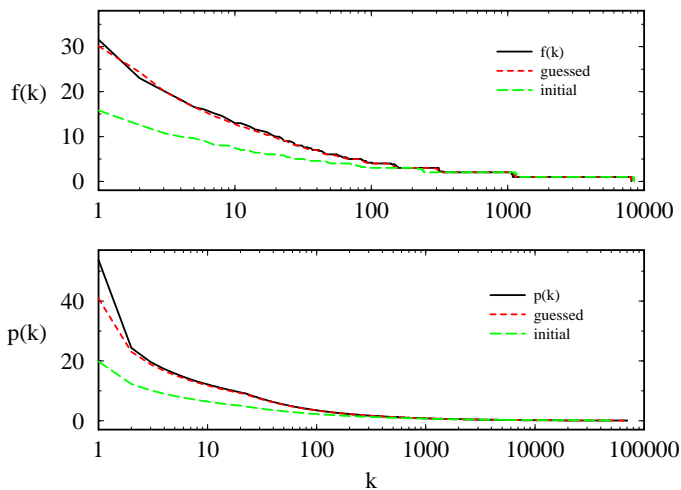,width=9cm,angle=270}}
\caption{The original and guessed probability distributions and the
Zipf--ordered frequencies for the distribution in eq.~(\ref{pade})
($N=10^4$).}
\label{test1}
\end{figure}

Now we try to guess a probability of a more complicated form
\begin{equation}
p(k)=\left\{ \begin{array}{r r l}
   \alpha\,(k - \epsilon)^{-\frac{1}{3}} &:& 1 \le k < \beta\\
   \phi\,k^{-\delta}                     &:& \beta \le k \le \gamma\\
    0                                    &:& k > \gamma~.
\end{array} \right.
\label{pade}
\end{equation}
(As we will show below this function approximates the probability
distribution of the words in an English text.) The variables $\alpha$
and $\phi$ can be eliminated due to normalization and continuity
condition. The test sample of size
$N=10^4$ was generated using $\epsilon=0.9$, $\beta=22$, $\delta
=0.64$ and $\gamma=70,000$. After the optimization we guess the
parameters $\epsilon^*=0.79$, $\beta^*=21.9$, $\delta^*=0.63$ and
$\gamma^*=65,000$. Fig.~\ref{test1} shows the original and the guessed
probability distributions and the Zipf--ordered frequencies for both
cases. The guessed entropy $H_{guess}=10.5053$ approximates the exact
value $H=10.5397$ very well while the observed entropy
$H_{obs}=8.8554$ shows a clear deviation from the correct value.
\subsection{Words in an English text}
\begin{figure}[ht]
\centerline{\psfig{figure=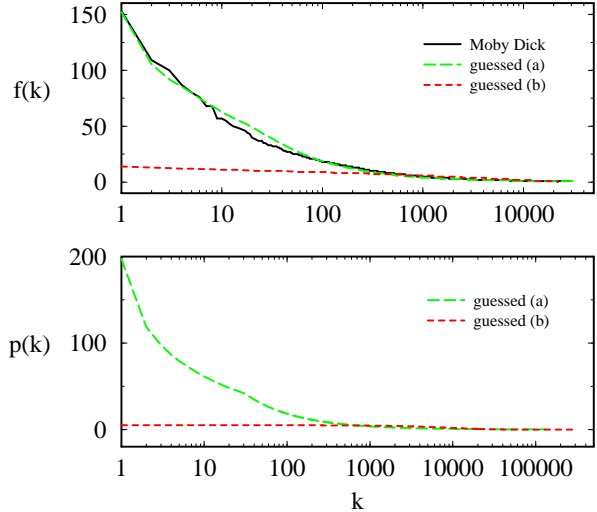,width=9cm,angle=270}} 
\caption{Zipf--ordered frequencies of words of length $n=6$ in ``Moby
 Dick''. The curves guessed (a) and (b) (top) display the frequency
 distributions which have been reproduced using the guessed
 probability distributions in the bottom figure according to
 eq.~(\ref{pade}) (a) and eq.~(\ref{simple}) (b).}
\label{beispiel}
\end{figure}

With the ansatz (\ref{pade}) we tried to guess the probability
distribution of the words of different length $n$ in the text ``Moby
Dick'' by H.~Melville~\cite{moby}. The text was mapped to an alphabet
of $\lambda=32$ letters as described in~\cite{ebpoe}. Depending on
overlapping or non--overlapping counting of the words we expect
different results. We note that overlapping counting is statistically
not correct since the elements of the sample are not statistically
independent, however, only overlapping counting yields enough words to
get somehow reasonable results for the observed entropy. We will show
that our method works in both cases, overlapping and non--overlapping.
Fig.~\ref{beispiel} shows the ordered frequencies of $N=5\cdot10^4$
words of the length $n=6$. The optimized distribution eq.~(\ref{pade})
reproduces the original frequency distribution (Moby Dick) with
satisfying accuracy.

\begin{figure}[ht]
\centerline{\psfig{figure=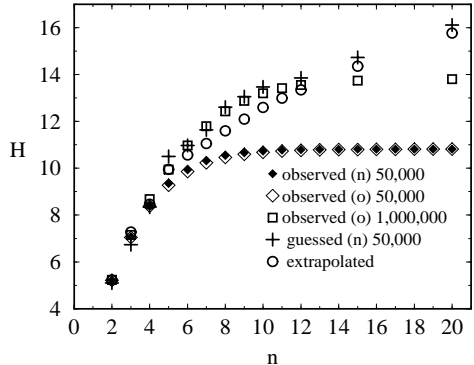,width=7cm,angle=0}}
\caption{\em Observed entropy $H^{(n)}_{obs}$ and guessed entropy
  $H^{(n)}_{guess}$ for the text ``Moby Dick'' ($N=5\cdot10^4$)
  over the word length $n$. The circles $\circ$ display the results of
  an extrapolation method described in [6] and the boxes $\Box$ show
  the observed entropy of the text using $N=10^6$ words. (o) denotes
  overlapping and (n) non-overlapping counting.}
\label{entropies}
\end{figure}

Using the ansatz~(\ref{pade}) we found $\epsilon^*=0.73$,
$\beta^*=31$, $\delta^*=0.70$ and $\gamma^*=129,890$. This calculation
was carried out for various word lengths $n$. Fig.~\ref{entropies}
shows the entropies $H^{(n)}_{obs}$ and $H^{(n)}_{guess}$ according to
eqs.~(\ref{hdefinition1},\ref{hdefinition}) as a function of $n$.
All results obtained have been derived from a set of $5\cdot 10^4$
non-overlapping words taken from the text of length $L=10^6$.  When we
count overlapping words, we find surprisingly that the entropy is
quite insensitive (see curves using filled and empty diamonds in
fig.~\ref{entropies}). The rather difficult problem of overlapping or
non--overlapping counting will be addressed in detail
in~\cite{EbelingPoeschel}. Since the exact probability distribution
for the words in ``Moby Dick'' is unknown we compare the guessed
entropy (crosses) with the observed entropy (empty diamonds:
overlapping counting, full diamonds: non-overlapping counting) and an
estimation of the entropy using an extrapolation method
(see~\cite{epa}), all based on the same set of data ($N = 5\cdot
10^4$), and with the observed entropy based on a twenty times larger
set of data (boxes: overlapping counting). As expected, for longer
word length $n$ the observed values $H^{(n)}_{obs}$ underestimate the
entropy. For small $n$ they are reliable due to the reliable
statistics. The guessed entropy $H_{guess}^{(n)}$ agrees for small $n$
with the observed entropy and for large $n$ with the extrapolated
values.

The form of the guessed theoretical distribution
$p(k,\alpha,\beta,\gamma,\dots)$ is arbitrary as long as it is
a normalized monotonically decreasing function (Zipf-order).
Suppose that one has no information about the mechanism which
generated a given sample. Then one has to find the functional form
of the guessed distribution which is most appropriate to a given
problem, i.e. in the ideal case the guessed distribution contains the
real probability distribution as special case without being too
complicated. An ansatz $p(k,\alpha,\beta,\gamma,\dots)$ is suited if
the optimized guessed probability distribution reproduces the
frequency distribution of the original sample with satisfactory
accuracy. 

The ansatz~(\ref{pade}) looks rather artificial: in fact we tried
several forms of the guessed probability distribution and the one proposed
in eq.~(\ref{pade}) turned out to be the best of them. None of the others
reproduces the frequencies sufficiently correct. For demonstration we
assume the function
\begin{equation}
p(k,\alpha,\beta)=\left\{
\begin{array}{r r l}
  \alpha\cdot\exp (-\beta\,k) &:& k\le \gamma\\
  0                          &:& k > \gamma
\end{array} \right.
\label{simple}
\end{equation}
with the normalization $\gamma=-\beta^{-1} \log\left(1- \beta /
\alpha\right)$. The optimized function is drawn in
figure~\ref{beispiel} ({\em guessed (b)}). We find that the frequency
distribution reproduced from this function differs much more from the
original frequency distribution (Moby Dick) than that of the guess
according to eq.~(\ref{pade}).

Admittedly any similar ansatz showing a well pronounced peak for low
ranks {\em (frequent words)}, a long plateau with slow decrease {\em
(standard words)} and a long tail {\em (seldom words)}, could give reliable
results as well. Anyhow there is no wide choice for the parametric
form of the probability distribution. Eq.~(\ref{pade}) belongs to the
class of distributions fulfilling this {\em three--region criterion}.
For a more detailed discussion of the statistics of words see
e.g.~\cite{ApostolicoGalil} and many references therein.
\section{Discussion}
The problem addressed in this paper was to find the rank ordered
probability distribution from the given frequency distribution of a
finite sample. For finite samples (Bernoulli sequence and English
text) we have shown that the calculation of the entropy using the
relative frequencies instead of the (unknown) probabilities yields 
wrong results.

We could show that the proposed algorithm is able to find the correct
parameters of a guessed probability distribution which reproduces the
statistical characteristics of a given symbolic sequence. The method
has been tested for samples generated by well defined
sources, i.e. by known probability distributions, and for an unknown
source, i.e. the word distribution of an English text. For the sample
sequences we have evidence that the algorithm yields reliable
results. The deviations of the entropy values from the correct values
are rather small and in all cases far better than the observed
entropies. For the unknown source ``Moby Dick'' we have no direct
possibility to check the quality of the method, however, the
calculated entropy values agree for small word lengths $n$ with the
observed entropy and for larger $n$ with the results of an
extrapolation method~\cite{epa}. In this sense both approaches support
each other. The proposed algorithm can be applied to the trajectories
of dynamic systems. Formally the trajectory of a discrete dynamics is
a text written in a certain language using $\lambda$ different
letters. The rank ordered distribution of sub-trajectories of length
$n$ belongs to the most important characteristics of a discrete
dynamic system. In this way we consider the analysis of English text
as an example for the analysis of a very complex dynamic system.

In many cases there is a principal limitation of the available data,
i.e. the available samples are small with respect to the needs of a
reliable statistics, and hence there is a principal limitation for the
calculation of the statistical properties using frequencies instead of
probabilities. For such cases using the proposed method one can
calculate values which could not be found so far. Given a finite set
of data the proposed method yields the most reliable values for the
Zipf--ordered distributions and the entropies which are presently
available. The method is not restricted to the calculation of the
entropy but all statistical properties which depend on the
Zipf--ordered probability distribution can be estimated using the
proposed algorithm.
\ack
{\samepage We thank T.~A\ss elmeyer, H.~J.~Herrmann and 
L.~Schimansky--Geier for
discussion and the {\em Project Gutenberg Etext, Illinois Benedictine
College, Lisle} for providing the ASCII--text of ``Moby Dick''.}


\begin{thebibliography}{99}
\bibitem{justice} J.~H.~Justice (Ed.),
{\em Maximum Entropy and Bayesian Methods in Applied Statistics},
Cambridge University Press (Cambridge, 1986).

\bibitem{herzel} H.~Herzel, {\em Syst.~Anal.~Mod.~Sim.} {\bf 5}, 435
(1988); P.~Grassberger, {\em Inf.~J.~Theor.~Phys.} {\bf 25}, 907 (1986);
{\em Phys.~Lett.~A} {\bf 128},369 (1988); {\em IEEE~Trans.~Inf.~Theo.} 
{\bf 35}, 669 (1989). 

\bibitem{schmitt} A.~Schmitt, H.~Herzel, and W.~Ebeling,
{\em Europhys.~Lett.} {\bf 23}, 303 (1993).

\bibitem{mcmillan} B.~McMillan, {\em Ann.~Math.~Statist.} {\bf 24},
196--216 (1953); A.~Khinchin, {\em Mathematical Foundation of
  Information Theory}, Dover (New York, 1967).

\bibitem{knuth} Donald~E.~Knuth, {\em The Art of Computer Programming}
Vol.~3, 506--570, Addison--Wesley (Reading, 1973); Robert Sedgwick,
{\em Algorithms}, Addison--Wesley (Reading, 1991).

\bibitem{epa} W.~Ebeling, T.~P\"oschel, and K.~Albrecht, {\em
Bifurcation \& Chaos} (in press).

\bibitem{recipes} W.~H.~Press, B.~P.~Flannery, S.~A.~Teukolsky, and
W.~T.~Vetterling, {\em Numerical Recipes}, Cambridge University Press
(Cambridge, 1987). 

\bibitem{moby} H.~Melville, {\em Moby Dick} (provided as ASCII--text by
Project Gutenberg Etext, Illinois Benedictine College, Lisle).

\bibitem{ebpoe} W.~Ebeling and T.~P\"oschel, {\em Europhys.~Lett.}
{\bf 26}, 241 (1994).

\bibitem{EbelingPoeschel} W.~Ebeling and T.~P\"oschel, {\em in
preparation}.

\bibitem{ApostolicoGalil} A.~Apostolico and Z.~Galil, {\em
Combinatorial Algorithms on Words}, Springer (Berlin, 19985). 

\end{thebibliography}
\end{document}